# Using Spatio-Temporal Dual-Stream Network with Self-Supervised Learning for Lung Tumor Classification on Radial Probe Endobronchial Ultrasound Video

Ching-Kai Lin[a,b,c,d], Chin-Wen Chen[a,*], Yun-Chien Cheng[a,*]

[a] *Department of Mechanical Engineering, College of Engineering, National Yang Ming Chiao Tung University, Hsin-Chu, Taiwan*

[b] *Department of Medicine, National Taiwan University Cancer Center, Taipei, Taiwan*

[c] *Department of Internal Medicine, National Taiwan University Hospital, Taipei, Taiwan*

[d] *Department of Internal Medicine, National Taiwan University Hsin-Chu Hospital, Hsin-Chu, Taiwan*

*<u>Corresponding author e-mail</u>: yccheng@nycu.edu.tw, wayne970331@gmail.com

## Abstract

The purpose of this study is to develop a computer-aided diagnosis system for classifying benign and malignant lung lesions, and to assist physicians in real-time analysis of radial probe endobronchial ultrasound (EBUS) videos. During the biopsy process of lung cancer, physicians use real-time ultrasound images to find suitable lesion locations for sampling. However, most of these images are difficult to classify and contain a lot of noise. Previous studies have employed 2D convolutional neural networks to effectively differentiate between benign and malignant lung lesions, but doctors still need to manually select good-quality images, which can result in additional labor costs. In addition, the 2D neural network has no ability to capture the temporal information of the ultrasound video, so it is difficult to obtain the relationship between the features of the continuous images. This study designs an automatic diagnosis system based on a 3D neural network, uses the SlowFast architecture as the backbone to fuse temporal and spatial features, and uses the SwAV method of contrastive learning to enhance the noise robustness of the model. The method we propose includes the following advantages, such as (1) using clinical ultrasound films as model input, thereby reducing the need for high-quality image selection by physicians, (2) high-accuracy classification of benign and malignant lung lesions can assist doctors in clinical diagnosis and reduce the time and risk of surgery, and (3) the capability to classify well even in the presence of significant image noise. The AUC, accuracy, precision, recall and specificity of our proposed method on the validation set reached 0.87, 83.87%, 86.96%, 90.91% and 66.67%, respectively. The results have verified the importance of incorporating temporal information and the effectiveness of using the method of contrastive learning on feature extraction.

## 1. Introduction

### 1.1 Background

Cancer poses a substantial public health challenge across the globe and ranks as the second highest cause of mortality in the United States in 2022 [1]. Lung cancer stands out as the most lethal form, underlining the vital importance of precise detection and proper therapy. Diagnosing peripheral pulmonary lesions (PPL) is a crucial step in devising appropriate treatment plans. Endobronchial ultrasound-guided transbronchial biopsy (EBUS-TBB) surgery has been widely applied in the examination of pulmonary lesions, with numerous studies confirming the reliability and safety of the technique [2-4]. The EBUS-TBB process can be divided into the following steps: (1) insertion of the ultrasound probe through the trachea into the pulmonary lesion area, (2) performing diagnostic endoscopic ultrasound to identify suitable regions for sampling, (3) conducting tissue biopsy when the physician observes malignant features in the lesion area, (4) staining the obtained sample with a smear technique, (5) performing cytopathological examination to preliminarily determine the benign or malignant nature of the lesion, (6) sending the sample for pathological analysis when there is sufficient specimen for evaluation. If only benign cells are identified, the process returns to step two and repeats the sampling procedure, as illustrated in Figure 1.

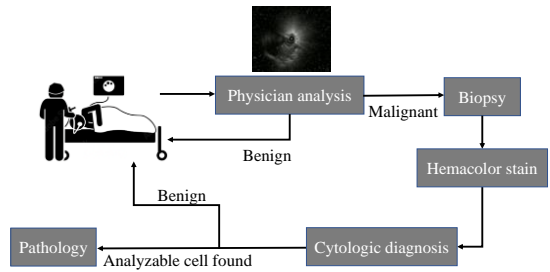

Figure 1. EBUS-TBB procedures.

During the EBUS-TBB sampling process, ultrasound imaging not only determines the location of the lesion but also analyzes its internal structure [5, 6]. Previous literature has used EBUS ultrasound images to differentiate between benign and malignant lesions, such as continuous margin, dotted or mottled air bronchogram, and heterogeneous echogenicity within the lesion [7, 8]. However, interpreting ultrasound image features for physicians performing bronchoscopy examinations requires considerable experience. Insufficient familiarity with these characteristics may increase the time spent searching for lesions, raise the probability of misjudgment, and elevate the risk of accidents during the surgical procedure. Consequently, researchers aim to develop more convenient and accurate methods by employing automated computer-aided diagnostic systems to assist physicians in determining the benign or malignant nature of ultrasound images during the examination process.

### 1.2 Literature Review

*1.2.1 CNN based approaches in medical ultrasound image classification*

In recent years, the trend has shifted from manual interpretation to computer-aided diagnosis. Many studies have attempted to use 2D convolutional neural networks (CNNs) for ultrasound image classification tasks, such as breast and thyroid ultrasound [9-12]. These studies have distinguished the benign and malignant lesions through CNNs, achieving results comparable to professional physicians and confirming the effectiveness of neural networks in interpreting medical ultrasound images. However, the 2D structure still has limitations and is difficult to apply in clinical diagnosis. First, ultrasound images are prone to noise, making them difficult to interpret. To obtain images suitable for discrimination, many tasks require manual selection and annotation of data, increasing the demand for manpower and difficulty in use. Second, ultrasound during sampling is real-time video, and the characteristics of lesions change over time. It is difficult to see all the features of a lesion from a single image, and the 2D structure cannot obtain the relationship between adjacent images' features, thus losing contextual information in the time direction.

Therefore, some studies have investigated the performance of three-dimensional convolutional neural networks (3D CNNs) in ultrasound video recognition [13, 14], such as echocardiography and fetal ultrasound, achieving better performance than 2D models in these tasks. For ultrasound video, the main difference between 3D and 2D lies in whether they can handle time-direction information. The input of 3D CNN considers multiple images at different time points, thus providing more information for judgment.

*1.2.2 CNN based approaches in EBUS-TBB tumor classification*

In terms of EBUS-TBB benign and malignant assessment, Chen's team proposed a CaffeNet-based architecture combined with support vector machines (SVMs) for analyzing the benign and malignant lesions in EBUS images [15], achieving an accuracy of 85.4% and an AUC of 0.87 in their dataset, establishing the ability of the CNN architecture to analyze EBUS image malignancy. Additionally, Takamasa's team proposed a deep convolutional network model, analyzed and verified the model's ability to interpret lesion ultrasound images, and compared the results of multiple professional physicians. Ultimately, the model's accuracy reached 83.3%, higher than the 73.8% and 66.7% of two experts, confirming that computer-aided interpretation can surpass professional physician diagnoses in this task. However, the datasets used in these studies were manually selected by professional physicians, and the 2D structure still has many limitations, leading to difficulties in clinical application.

To overcome these limitations and improve the clinical applicability of computer-aided diagnosis systems for EBUS-TBB, future research could focus on developing more advanced 3D CNN models that can better capture the temporal context and spatial relationships between adjacent images. The use of a 3D CNN model has the potential to eliminate the need for physicians to manually select each frame of an image in a video.

In conclusion, the adoption of computer-aided diagnosis systems, particularly those based on 3D CNNs, has the potential to revolutionize the interpretation of EBUS-TBB images, improving accuracy and efficiency in detecting benign and malignant lesions. Further research and development in this area will pave the way for more advanced, user-friendly, and clinically applicable systems that can assist physicians in making better diagnostic decisions and ultimately improve patient care.

## 1.3 Motivation

In light of the aforementioned studies, it is evident that the CNN architecture has been widely used and achieved tangible results in the field of ultrasound image interpretation. However, there is still room for improvement in clinical applications. Firstly, we aim to use ultrasound recordings as model inputs to reduce the labor cost required for physicians to manually select images. Secondly, we plan to design models using a 3D architecture to obtain the temporal context information, thus reducing reliance on individual images. Thirdly, we hope to minimize the interference of noise on the model's predictive results and enhance classification capabilities under different environments. This study aims to establish an automated computer-aided interpretation system to assist physicians in diagnosing the benignity and malignancy of lung lesions during EBUS-TBB, thereby improving diagnostic accuracy and reducing surgical risks. We propose an architecture that effectively utilizes video information by integrating temporal and spatial direction features, while also mitigating the impact of noise through contrast learning. We hope to achieve a higher classification accuracy than existing methods.

## 2. Material and Methods

### 2.1 EBUS-TBB Dataset

Ultrasound recordings of EBUS-TBB were retrospectively collected from the National Taiwan University Cancer Center between November 2019 and April 2021. The dataset comprised 309 cases, which were split into training and validation sets at an 8:2 ratio. The training set included 247 cases, consisting of 175 malignant and 72 benign cases, while the remaining 62 cases formed the validation set, containing 44 malignant and 18 benign cases. This study received approval from the Institutional Review Board of the National Taiwan University Cancer Center (IRB #202207064RINB).

## 2.2 Experimental Procedure

As illustrated in Figure 2, the study began by collecting EBUS-TBB recordings and annotating the occurrence times of lesions. Subsequently, the dataset was divided, preprocessed, and trained using SlowFast and SwAV models. In this section, each step of the process will be discussed in detail.

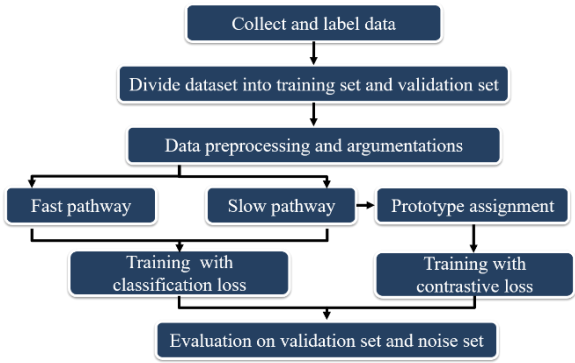

Figure 2. Experiment process.

## 2.3 Data Preprocessing

Each patient includes multiple ultrasound recordings of less than 10 minutes in length. The videos are sampled at a rate of one frame every 0.4 seconds, with every 8 frames stacked into a single three-dimensional image as the input for the model, referred to as a clip. To prevent the loss of temporal information caused by splitting adjacent images into different clips, there is a 50% overlap between neighboring clips. As a result, each video clip contains 3.2 seconds of information with a 1.6-second time difference between every two adjacent clips, as shown in Figure 3.

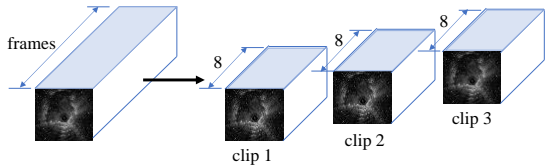

Figure 3. Video sampling process.

The images are then cropped to 960×960 pixels and downsampled to 224×224 pixels to reduce computation. To avoid overfitting of the training data, horizontal flipping with a 50% probability is employed as a data augmentation method. Additionally, the Noise CutMix method is used in some tests for SwAV model training and noise impact verification.

## 2.4 Noise CutMix

In this study, we utilized the CutMix method to convert original images into images containing noise information, as depicted in Figure 4. CutMix is a data augmentation method that replaces a random region of the original image with another image from the dataset. In our case, we substitute a specific position of the original image with a noise image, hence the name "Noise CutMix". Since the main source of noise in ultrasound is air echo, which appears as a fan shape in EBUS-TBB, we simulate realistic noise by randomly adding 1/4 circle noise images in the four corners of the original image. The noise images were selected from 237 noise images in the dataset.

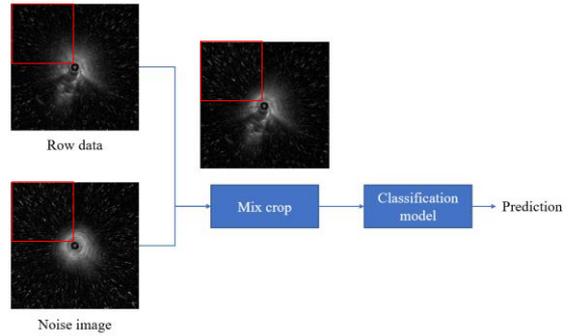

Figure 4. Noise CutMix process.

## 2.5 Model Overview

The training of the model can be mainly divided into two parts: SlowFast and SwAV, as shown in Figure 5. First, the video undergoes preprocessing, using K types of data augmentation methods and expanding the data volume by K times. Next, the SlowFast model extracts and fuses features from both the temporal and spatial dimensions. Finally, supervised and unsupervised losses are calculated separately. The classification loss of the prediction network is supervised and is used to enhance the model's ability to discern between benign and malignant lesions. The contrastive loss of SwAV is unsupervised, allowing images with the same content but processed using different data augmentation methods to have closer representations in the feature space, reducing the impact of noise and promoting the clustering tendency of features.

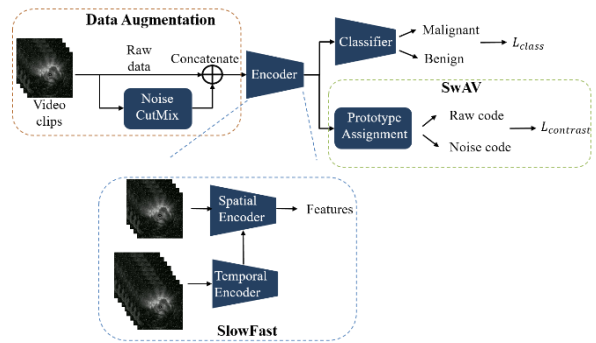

Figure 5. Model overview.

## 2.6 SlowFast

In video recognition, 3D networks can extract temporal information better than 2D architectures and have improved interpretation capabilities for moving objects. However, the required computational resources also increase accordingly. Although increasing the sampling frequency and the number of input frames can provide more temporal information, the computational cost grows in proportion. To efficiently obtain more temporal information, Feichtenhofer's team proposed the SlowFast dual-stream network architecture [16]. The model is divided into Slow and Fast pathways. The Slow pathway is mainly used to extract static features in the spatial dimension, so a smaller number of input frames and a higher number of channels are utilized. In this case, 3D ResNet50 is chosen as the backbone network. The Fast pathway is used to extract dynamic features in the temporal dimension, so more input frames are needed, and the number of channels is reduced to alleviate computational burden. Here, we use a sampling frequency four times higher, with 32 frames as the input for the Fast pathway. The feature maps from each layer of the Fast pathway are reshaped using 3D convolution and added to the corresponding layer of the Slow pathway to fuse spatial and temporal features.

## 2.7 SwAV

Contrastive learning is a relatively new unsupervised learning method that enhances the model's clustering ability by dividing the data into positive and negative samples. It aims to bring positive samples, which are similar to each other, closer in the feature space while increasing the distance between dissimilar negative samples. SwAV is a contrastive learning method that does not use negative samples [17]. It involves swapping predictions of the same image transformed differently. Instead of directly comparing the features extracted by the model as in other contrastive learning methods, SwAV calculates the similarity between features and K cluster centers (prototypes) to generate soft-label codes. These codes are then used as labels for swapping predictions among themselves.

## 3. Results
### 3.1 Evaluation Methods

To verify the effectiveness of temporal information in distinguishing between benign and malignant lung lesions, various 2D models such as 2D ResNet50, GoogLeNet, VGG19, and ViT, and 3D models like 3D ResNet50 and SlowFast will be trained and compared. The validation set will be augmented with 50% noise images using the Noise CutMix method, resulting in the "noise dataset". This dataset will be utilized to evaluate the model's robustness to noise and its ability to maintain performance in the presence of noisy data. Additionally, the average distribution of benign and malignant cases' codes in the clustering space (prototypes) will be compared to examine the effectiveness of contrastive learning. Since the video length of each patient is different, the average of the segment predictions will be used as the case prediction. Cases with a prediction value greater than or equal to 0.5 will be considered malignant. AUC, accuracy, precision, recall, and specificity will be used as evaluation metrics for the validation set.

### 3.2 Evaluation of Different Models in Validation Set

As shown in Table 1, the SlowFast+SwAV model achieved the highest classification scores, with AUC, accuracy, recall, and specificity reaching 0.87, 0.84, 0.91, and 0.67, respectively. Among the 2D models, VGG19 performed best with an AUC of 0.85, even surpassing 3D ResNet50's 0.848. The results show that 2D models are less effective in accurately classifying benign cases, as their specificity was lower compared to 3D models. Additionally, the SlowFast model demonstrated significantly improved overall scores compared to the 3D ResNet50 model, suggesting that incorporating additional temporal information is beneficial for our dataset. Furthermore, the results demonstrate that contrastive learning can further enhance the classification performance of the model.

### 3.3 Evaluation of Different Models in Noise Dataset

As shown in Table 2, the SlowFast+SwAV model maintained the highest classification scores, with AUC, accuracy, recall, and specificity reaching 0.872, 0.82, 0.91, and 0.61, respectively, showing only slight differences compared to the noise-free scenario. Among the 2D models, VGG19 and GoogLeNet experienced more significant declines, while the scores of other models changed only slightly.

### 3.4 The Impact of SwAV on Spatial Clustering

The results show that, without SwAV training, the feature mapping does not exhibit any clustering in the clustering space, regardless of whether the model undergoes classification loss training or not, as illustrated in Figures 6 and 7. However, the clustering effect of features can be effectively improved by adding SwAV training, as shown in Figure 8. Compared to the SlowFast model without SwAV, the method with SwAV can better express the differences between benign and malignant cases. Figure 9 shows that even with the addition of noise in the dataset, the features can still be clearly clustered, indicating that the features trained with SwAV will not be significantly affected by noise in their distribution.

Table 1. Classification result in validation set.

| | Model | AUC | Accuracy | Recall | Specificity |
|---|---|---|---|---|---|
| **2D** | GoogLeNet | 0.81 | 0.71 | 0.83 | 0 |
| | 2D ResNet50 | 0.81 | 0.71 | **0.98** | 0.056 |
| | VGG19 | 0.85 | 0.77 | **0.98** | 0.27 |
| | ViT | 0.83 | 0.79 | **0.98** | 0.33 |
| **3D** | 3D ResNet50 | 0.848 | 0.74 | 0.88 | 0.39 |
| | SlowFast | 0.862 | 0.80 | 0.95 | 0.44 |
| | SlowFast+SwAV | **0.87** | **0.84** | 0.91 | **0.67** |

Table 2. Classification result in noise dataset.

| | Model | AUC | Accuracy | Recall | Specificity |
|---|---|---|---|---|---|
| **2D** | GoogLeNet | 0.74 (-0.07) | 0.71 (+0) | **1.0** (+0.17) | 0 (+0) |
| | 2D ResNet50 | 0.82 (+0.01) | 0.71 (+0) | **1.0** (+0.02) | 0 (-0.056) |
| | VGG19 | 0.83 (-0.02) | 0.71 (-0.06) | 0.83 (-0.15) | 0.056 (-0.21) |
| | ViT | 0.83 (+0) | 0.77 (-0.05) | 0.98 (+0) | 0.028 (-0.3) |
| **3D** | 3D ResNet50 | 0.851 (+0.003) | 0.76 (+0.02) | 0.88 (+0) | 0.44 (+0.05) |
| | SlowFast | 0.86 (-0.002) | 0.73 (-0.07) | **1.0** (+0.05) | 0.056 (-0.38) |
| | SlowFast+SwAV | **0.872** (+0.002) | 0.82 (-0.02) | 0.91 (+0) | **0.61** (-0.06) |

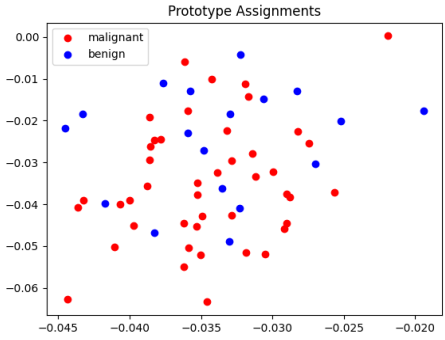

Figure 6 Untrained SlowFast model's prototype assignment.

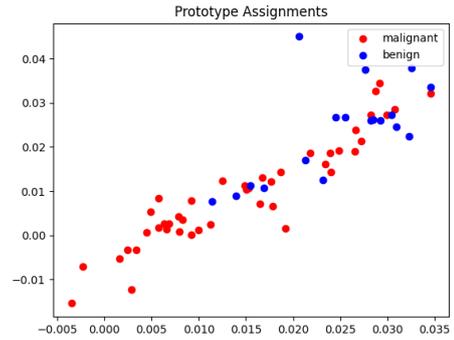

Figure 8 Trained SlowFast+SwAV model's prototype assignment.

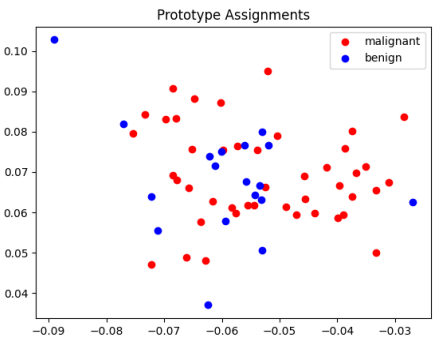

Figure 7 Trained SlowFast model's prototype assignment.

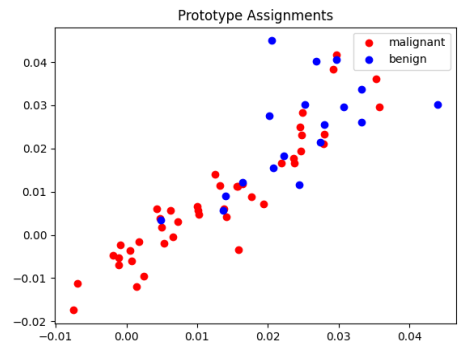

Figure 9 Trained SlowFast+SwAV model's prototype assignment in noise dataset

## 4. Conclusion

This study developed a model that combines SlowFast, which includes fused spatiotemporal information, and SwAV, which utilizes contrastive learning. The model was applied to interpret lung lesions in bronchoscopic ultrasound images and demonstrated that the inclusion of temporal information improved the model's performance. Additionally, contrastive learning was found to enhance the clustering capability of features. The SlowFast+SwAV architecture achieved the highest AUC of 0.87 and outperformed human judgment in various indicators.

In the future, the diagnostic model proposed in this study could serve as a basic guideline for physicians in determining the benign or malignant nature of lesions during EBUS-TBB, thereby increasing their confidence and accuracy in diagnosis and reducing surgical time and risk. To enhance the model's performance, we intend to explore using a transformer-based architecture for prediction and test other forms of contrastive learning while conducting more comprehensive ablation tests.